\documentclass[prb,twocolumn,aps, amsmath, amssymb,showpacs]{revtex4-1}
\usepackage{epsfig}
\usepackage{graphicx}

\begin{document}

\title{Electronic Structure and Properties of SrAlGe and BaAlGe}

\author{S. J. Youn$^{1,2}$ and A. J. Freeman$^{1}$}

\address{$^1$ Department of Physics Education and Research Institute of Natural Science, Gyeongsang National University, Jinju 660-701, Korea}
\address{$^2$ Department of Physics and Astronomy, Northwestern University, Evanston, Illinois, 60208-3112, USA}


\date{\today}
\begin{abstract}
The electronic structures of BaAlGe and SrAlGe which are superconductors with hexagonal honeycomb layers have been studied by using a first principles method. Energy bands, Fermi surafces, and density of states are presented.
The two materials have topologically different Fermi surfaces.
BaAlGe has two Fermi surfaces: One has a three dimensional spinning-top-like shape and the other has a cylindrical shape with two dimensional character. SrAlGe has only one connected Fermi surface.
Two gap superconductivity for BaAlGe is suggested from the inherently different character of the two Fermi surfaces.
The higher $T_c$ of SrAlGe than BaAlGe is related to the difference in both the topology of the Fermi surface and the band dispersions along the $z$ direction.
\end{abstract}
\maketitle


\section{Introduction}
\label{}



The discovery of superconductivity in MgB$_2$ with $T_c$=39K\cite{nagamatsu} has attracted continuing attention  due to interests in both fundamental research\cite{moshchalkov} and practical applications.\cite{collings}
It crystalizes in the AlB$_2$ type structure with a space group of $P6/mmm$ ($D^1_{6h}$, No.191).\cite{hoffmann}
Numerous efforts to find a novel superconductor with higher $T_c$ in the AlB$_2$ family have been unsuccessful so far.
There are eight valence electrons in a unit cell which belong to two groups according to the symmetry of wave functions: $\sigma$ bands originate from B $s$, $p_{x,y}$ orbitals and form a strong honeycomb network of boron atoms while $\pi$ bands originate from B $p_z$ orbitals and show three dimensional conductivity.
Electron-phonon coupling between $E_{2g}$ in-plane phonon mode and in-plane $\sigma$ electrons in a boron layer is the main superconductivity mechanism.\cite{an}
Due to the two inherently different bands, MgB$_2$ has two different superconducting energy gaps.\cite{kortus,choi}

Related to this, studies of superconductivity in the ternary silicides, $MA$Si($M$=Ca,Sr,Ba, $A$=Al,Ga), were triggered by CaSi$_2$ which becomes a superconductor ($T_c$=14K) at high pressure above 16 GPa, while it is a semimetal at ambient pressure.\cite{sanfilippo}
It undergoes a phase transition with pressure from a trigonal structure with corrugated Si layers to the AlB$_2$ structure.
$MA$Si has nine valence electrons with one more electron than MgB$_2$.\cite{fahy}
CaAlSi has the highest $T_c$=7.8K among the ternary silicides.\cite{imai2002}
However, it was found that the structure consists of 6 slabs, 6$H$-CaAlSi, whlie the $T_c$ of one slab, 1$H$-CaAlSi, is 6.5K.\cite{kuroiwa2006}
The ternary silicides are metallic with one partially filled $\pi^*$ band originated from antibonding $p_z$ orbitals while the $\sigma$ bands are fully occupied;\cite{huang}
they exhibit electron-like conductivity which is different from the hole-like conductivity of MgB$_2$.\cite{lorenz}
The $d$ electrons in $M$ metal atoms, for example, of Ca 3$d_{xy}$ and $d_{z^2}$ origin, play an important role which is different from MgB$_2$ with only $s$ and $p$ electrons;\cite{mazin,giantomassi}
$M$ metal atoms also play an active role in determining the lattice dynamics.\cite{huang}
It is reported that the electron-phonon interaction mediated by a soft phonon mode vibrating along the $z$-direction gives rise to BCS type superconductivity.\cite{kuroiwa2008}

\begin{figure}[bt]
\centering
\begin{tabular}{cc}
\epsfig{file=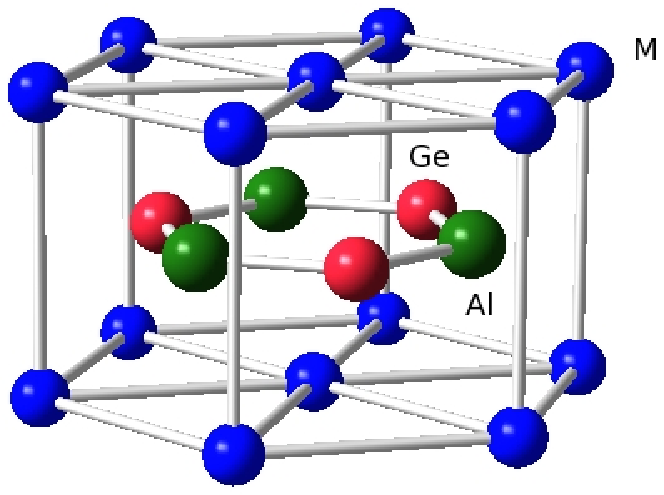,width=0.5\linewidth,clip=} &
\epsfig{file=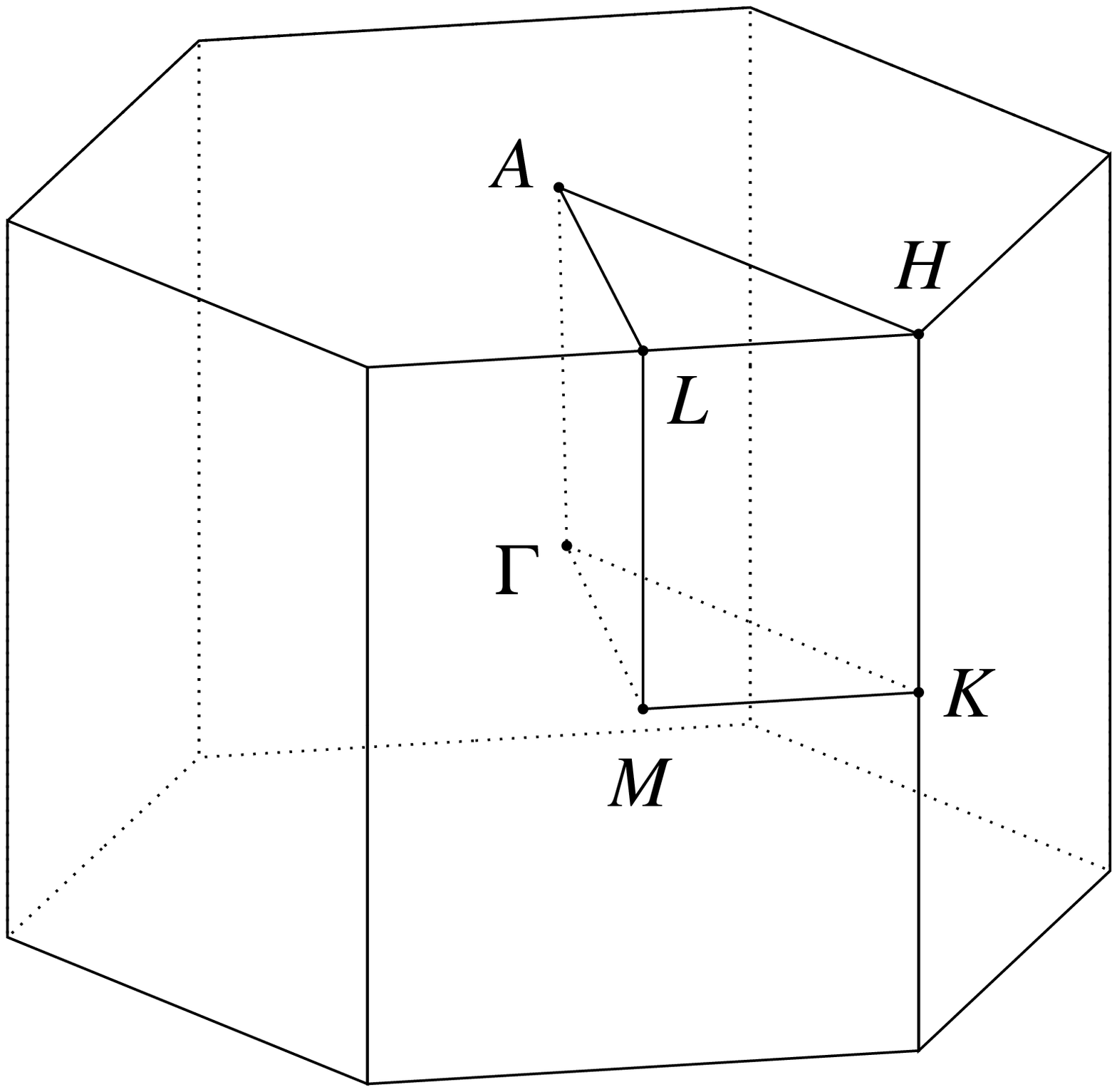,width=0.35\linewidth,clip=}\\
\mbox{\bf (a)} & \mbox{\bf (b)}
\end{tabular}
\caption{(a) crystal structure and (b) hexagonal Brillouin zone of $M$AlGe. The $M$ site in (a) represents a site for Ba or Sr. In (b), symbols are marked only around the irreducible Brillouin zone. \label{fig1}}
\end{figure}

Recently ternary superconductors in the AlB$_2$ family were extended to include germanides with nine valence electrons.\cite{evans}
Among them,
SrAlGe ($T_c$=6.7K) and BaAlGe ($T_c$=6.3K) are the two highest $T_c$ superconductors at ambient pressure.
Both SrAlGe and BaAlGe crystallize in a hexagonal structure of BaPtSb type with space group of $P\bar{6}m2$ ($D_{3h}^1$, No.187), as depicted in Fig. \ref{fig1}.
The crystal structure is similar to that of MgB$_2$ where the boron layers are replaced by AlGe layers.
There is a horizontal mirror symmetry $\sigma_h$ without inversion symmetry.
Due to the $\sigma_h$ symmetry, wave functions can have either even ($\sigma$ band) or odd parity ($\pi$ band) with respect to $\sigma_h$.

The exact understanding of the electronic structure is crucial to understand the mechanism of superconductivity.
Especially the character of the Fermi surface is important in determining the superconducting properties.
We report in this paper the electronic structure of SrAlGe and BaAlGe by using a first principles method.
Energy bands, Fermi surfaces, and densities of states will be presented in Sec.~\ref{sec:results}.
The different character between BaAlGe and SrAlGe will be addressed.

\section{Method}
First-principles calculations were performed using the
full-potential linearized augmented plane wave (FLAPW) method\cite{flapw,flapw2} in the local density approximation (LDA) for the exchange-correlation functional by Hedin and Lundqvist.\cite{hedin} Experimental lattice constants, $a$= 4.3043\AA~ and $c$= 4.7407\AA~ for SrAlGe and $a$= 4.3512\AA~ and $c$= 5.1401\AA~ for BaAlGe, are employed.\cite{evans}
$K_{max}$=3.7 and $l_{max}$=8 were used for both materials.

Muffin-tin radii are taken to be 2.6, 2.4 and 2.2 a.u for Ba(Sr), Al, and Ge, respectively.
Semicore electrons such as Sr $4p$ and Ge $3d$ are treated as valence electrons,
which are explicitly orthogonalized to the core states.\cite{explicit}
Although there is no inversion symmetry, time reversal symmetry plus the crystal symmetry
reduce the volume integration in $k$ space to 1/24 of the whole Brillouin zone.\cite{ramirez}
Summations in the Brillouin zone are done with a 11$\times$11$\times$11 mesh in the Monkhorst-Pack scheme during the self-consistent cycles,\cite{kpts}
while the density of states is obtained by the tetrahedron method with a 15$\times$15$\times$15 mesh.\cite{blochl}
In order to calculate Fermi velocity and plasma frequency at the Fermi level, eigenvalues from the self-consistent calculations
are fitted by a spline method over the whole Brillouin zone.\cite{spline}

\section{Results}
\label{sec:results}
The electronic structure of BaAlGe is presented first since
it is similar to the known electronic structure of isoelectronic ternary silicides.\cite{huang}
Then results for SrAlGe are presented with emphasis on the difference between the two materials.

\subsection{BaAlGe}
\begin{figure}[bt]
\epsfxsize=\hsize \center{\epsfbox{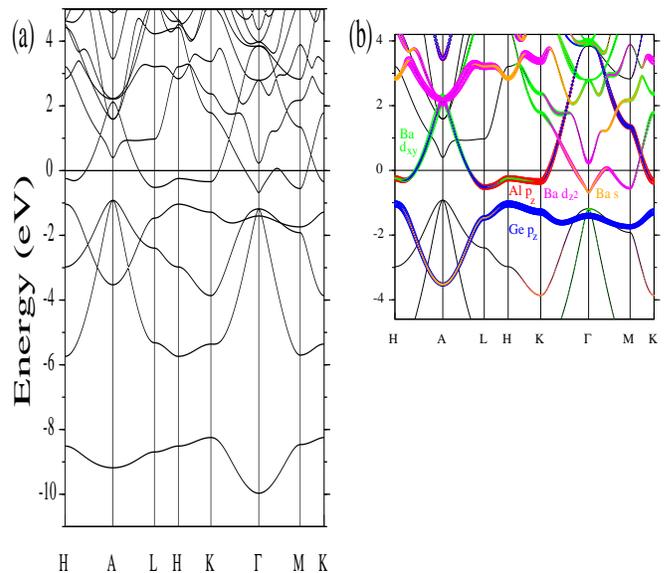}}
\caption{(a) Energy bands and (b) major orbital component of BaAlGe around $E_F$(set at 0). \label{fig2}}
\end{figure}

Figure~\ref{fig2}(a) depicts the energy bands along the symmetry lines of the hexagonal Brillouin zone shown in Fig. \ref{fig1}(b).
The first band around -9 eV comes from mostly the Ge 4$s$ orbital.
The 4.8 eV wide band from -5.8 eV to -1 eV is the $\sigma$ hybridized band between Al $s,p_{x,y}$ and Ge $p_{x,y}$ orbitals
where the minimum is at $A$ and the maximum is at $H$.
The $\sigma$ band width of MgB$_2$ is about 8 eV, which is much wider than that in BaAlGe.
The narrower band width of BaAlGe can be ascribed to weak covalent bonding between Al and Ge
since strong ionic bonding between Al and Ge is expected due to their large difference in electronegativity.\cite{pearson}

Figure~\ref{fig2}(b) exhibits band components around the Fermi level.
Only the fifth band crosses the Fermi level($E_F$) while the $\pi$ band of mostly Ge $p_z$ origin is fully occupied,
which is another manifestation of the large electronegativity of Ge.
The fifth band along $HK$ is of Al $p_z$ origin and displays little dispersion with a width of 0.10 eV
which gives rise to a peak in the density of states in Fig. \ref{fig4}.
The narrow band width in the $z$ direction is attributed to the weak inter-layer coupling of the Al $p_z$ orbitals in BaAlGe.
The fifth band exhibits different character depending on the location in the Brillouin zone;
Sr $d_{xy}$ character is found along $H-A-L$ while
Sr $d_{z^2}$ and $s$ character are found along $K-\Gamma-M$.
Around the $A$ point, $\pi$ and $\pi^*$ bands show parabolic behavior;
the $\pi$ bonding band is open upward while the $\pi^*$ antibonding band is open downward, as shown in Fig. \ref{fig2}(b).
Although the $\pi^*$ band also shows antibonding parabolic behavior around $\Gamma$, the $\pi$ band changes to a weak bonding character at $\Gamma$.
One can clearly see the $\pi$-$\pi^*$ splitting along $L-H-K$ in Fig. \ref{fig2}(b) because the AlGe honeycomb layer is made of different atoms.
When the honeycomb layer is made of the same atoms like MgB$_2$, the $\pi$ and $\pi^*$ bands are degenerate along $HK$.

\begin{figure}[bt]
\centering
\begin{tabular}{cc}
\epsfig{file=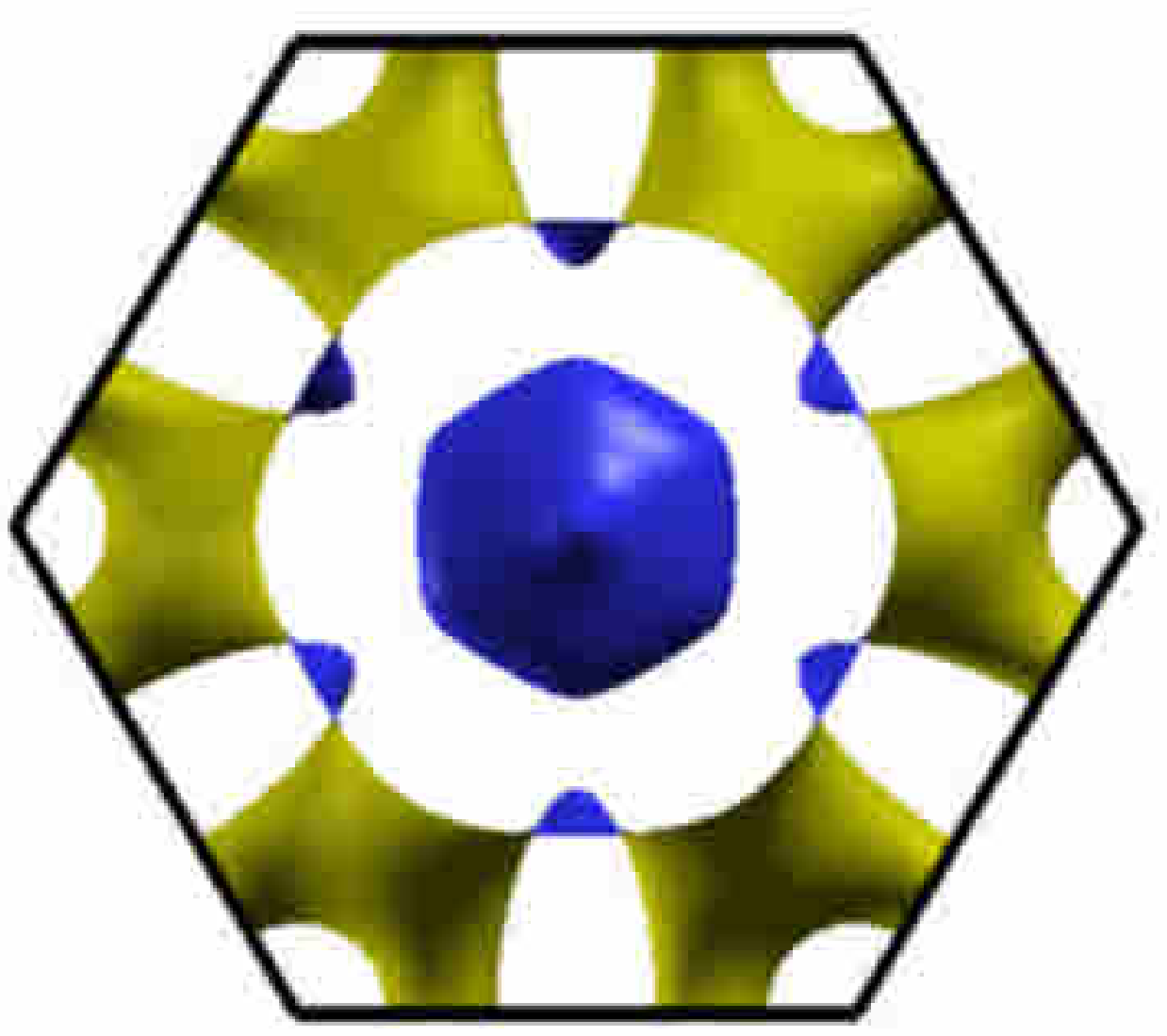,width=0.5\linewidth,clip=} &
\epsfig{file=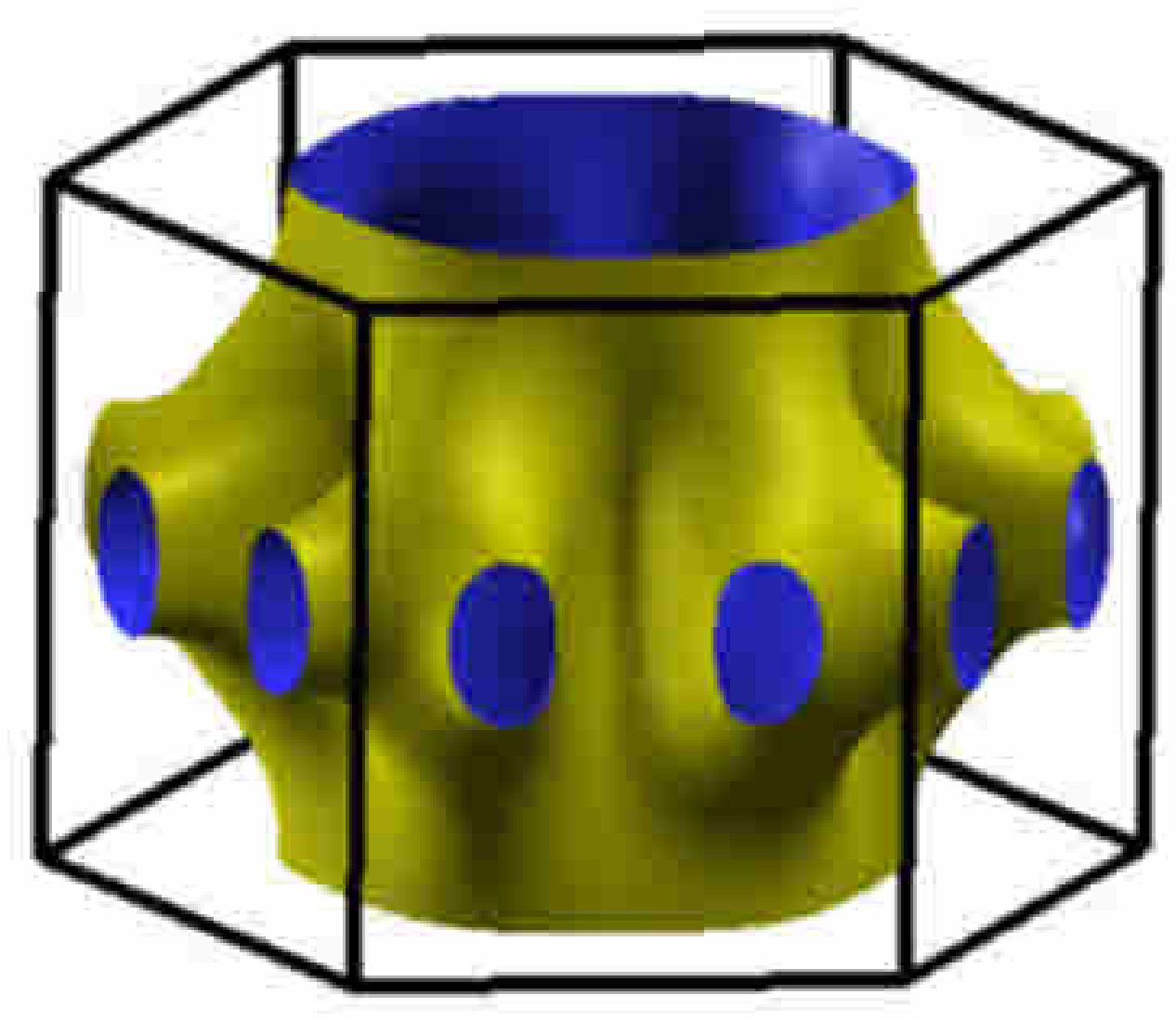,width=0.5\linewidth,clip=}\\
\mbox{\bf (a)} & \mbox{\bf (b)}\\
\epsfig{file=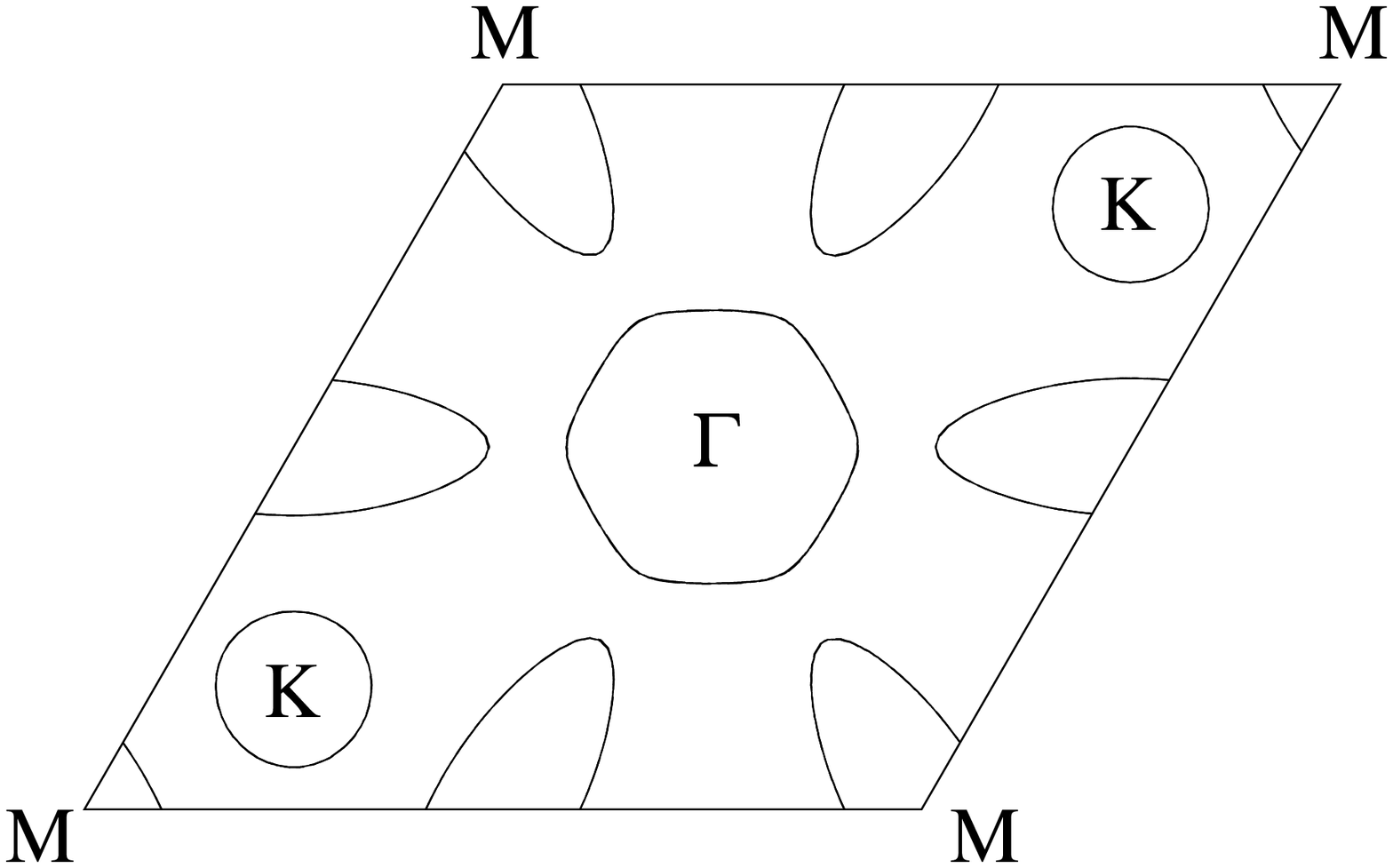,width=0.5\linewidth,clip=} &
\epsfig{file=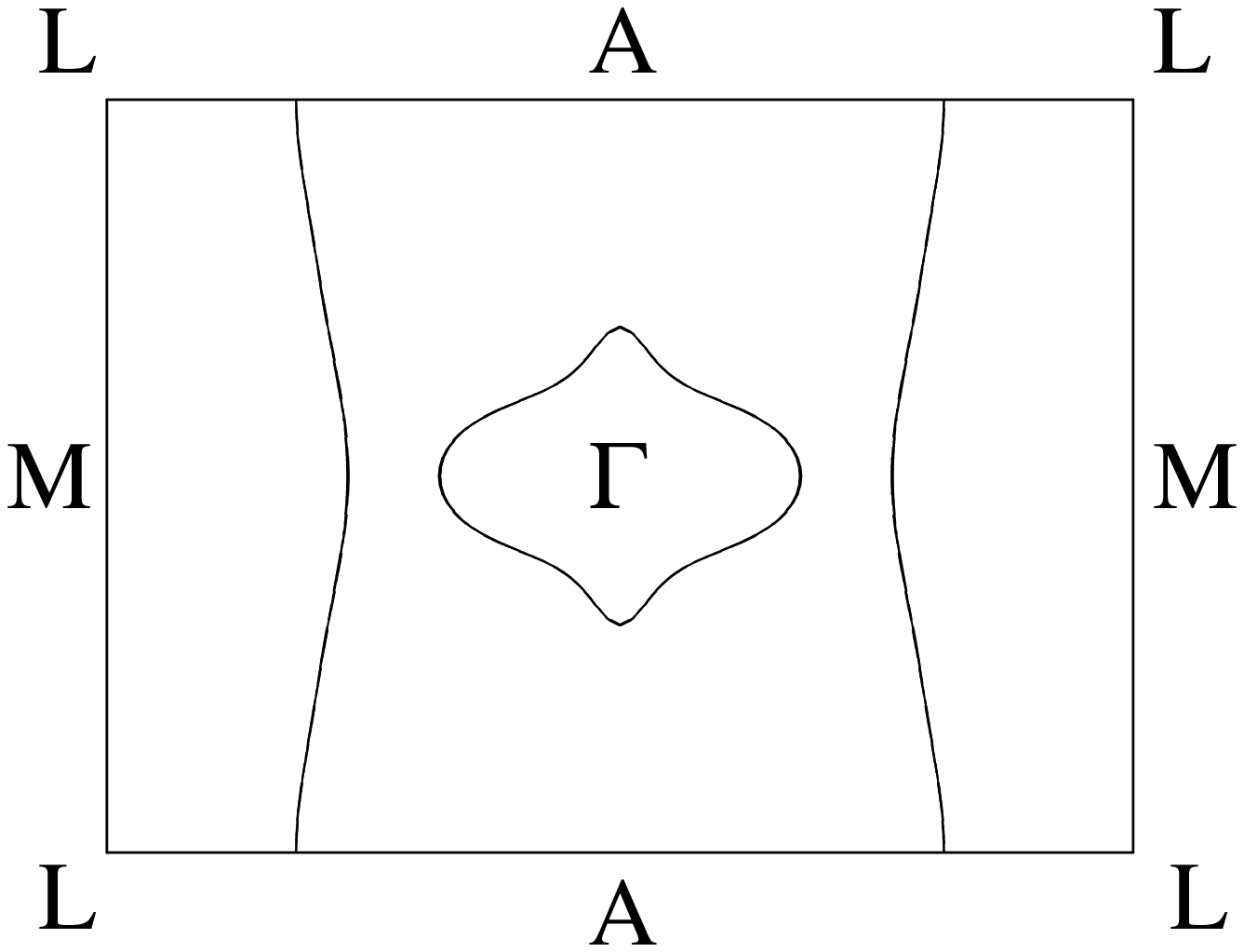,width=0.4\linewidth,clip=}\\
\mbox{\bf (c)} & \mbox{\bf (d)}
\end{tabular}
\caption{ Fermi surfaces of BaAlGe. (a) Top view and (b) side view. In (b), the body center of the hexagonal prism is the $\Gamma$ point. Blue and golden colors represent electron-like and hole-like surfaces, respectively. Cross-section in (c)(001) plane and (d)(100) plane.\label{fig3}}
\end{figure}

\begin{figure}[bt]
\epsfxsize=0.7\hsize \center{\epsfbox{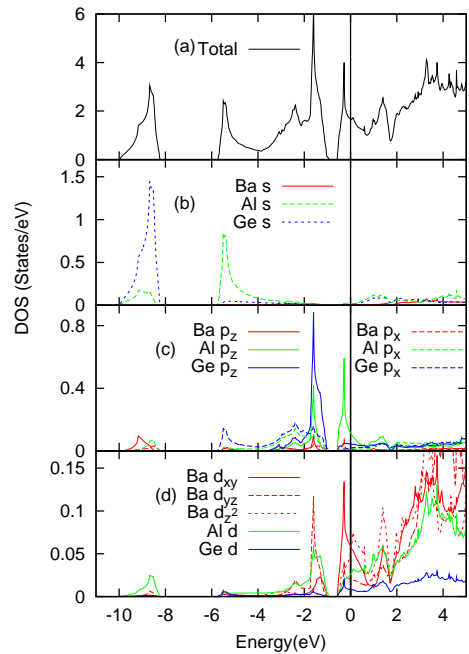}}
\caption{Total and partial density of states of BaAlGe.\label{fig4}}
\end{figure}

There are two separate Fermi surfaces as shown in Fig. \ref{fig3}: an electron-like inner pocket centered at $\Gamma$ with the shape of a spinning top and the hole-like cylindrical outer sheet.
The shapes of the Fermi surface can be seen easily in the cross-section of Fig. \ref{fig3}(d).
As shown in Fig. \ref{fig2}(b), the cylindrical sheet originates from Ba $d_{xy}$ and Al $p_z$ orbitals,
while the electron pocket is from both the Ba $d_{z^2}$ orbital and $s$ orbitals (Ba $s$, Al $s$, Ge $s$).
Since the symmetry of Ba $d_{z^2}$ is different from $p$ orbitals,\cite{mattheiss} it is nonbonding with other $p$ orbitals.
Thus the two Fermi surfaces can be called antibonding and nonbonding for the pocket and cylinder Fermi surfaces, respectively, from the character of component orbitals.
Cylindrical Fermi surfaces are connected to other cylindrical Fermi surfaces in neighboring Brillouin zones by twelve bridges around the $K$ points.
A similar topology of Fermi surfaces is also found in ternary silicides.\cite{giantomassi,kuroiwa2007}
The two disconnected Fermi surfaces are also manifested in the energy band of Fig. \ref{fig2} where the fifth band crosses the Fermi level twice along $\Gamma M$.
Two Fermi surfaces with different character may give rise to a two gap superconductor.
Actually, two superconducting gaps were suggested from optical measurements for an isoelectronic silicide, CaAlSi.\cite{lupi}
Note the circular hole around K which can be seen only in the cros-sectional plot of Fig. \ref{fig3}(c).

Fermi surface properties are calculated from the fitted energy bands by the spline method.\cite{spline}
In-plane and out-of-plane Fermi velocities are
$\langle v_{x,y}^2\rangle^{1/2}$=3.10$\times 10^7$ cm/s and $\langle v_z^2\rangle^{1/2}$=2.06$\times 10^7$ cm/s, respectively.
The anisotropy ratio of the conductivity is $\sigma_{xx}/\sigma_{zz}=\langle v_x^2\rangle/\langle v_z^2\rangle$=2.26
where an isotropic scattering rate is assumed;
the theoretical anisotropic ratio of MgB$_2$ is 1.06, which is much lower than that of BaAlGe.\cite{kortus}
The plasma frequencies are $\Omega_{x,y}$=3.83 eV and $\Omega_z$=2.55 eV.
The optical and transport properties are expected to be anisotropic between the $x$ and $z$ directions.
Since the two Fermi surfaces are separated, contributions from each surface can be calculated, which are summarized in Table~\ref{table1}.
Note that the Fermi velocity in the $z$ direction is greater than in the $x$ direction on the pocket Fermi surface,
which can be attributed to inter-layer coupling by hybridization between Ba $s, d_{z^2}$ and Al $s$, Ge $s$ orbitals.
Thus, interlayer coupling is made from non-bonding orbitals such as Ba $s, d_{z^2}$ and Al $s$, Ge $s$ orbitals, which can give rise to a strong electron-phonon coupling along the $z$ direction.

\begin{table}[b]
\caption{\label{table1}
Fermi surface properties of BaAlGe and SrAlGe: Density of states, $N(E_F)$, average velocities, $\langle v_x^2\rangle^{1/2}$,  $\langle v_z^2\rangle^{1/2}$, anisotropy ratio, $\langle v_x^2\rangle/\langle v_z^2\rangle$, and plasma frequencies, $\Omega_x$, $\Omega_z$ of which units are States/eV/spin, $10^7$cm/s, and eV, respectively.
}
\begin{ruledtabular}
\begin{tabular}{ccccccc}
  & $N(E_F)$& $\langle v_x^2\rangle^{1/2}$  & $\langle v_z^2\rangle^{1/2}$ &$\langle v_x^2\rangle/\langle v_z^2\rangle$   & $\Omega_x$  & $\Omega_z$ \\ \hline
BaAlGe & &&&&&\\
 pocket & 0.17 & 2.27 & 3.75 & 0.37 & 0.90 & 1.48 \\
 cylinder & 1.46 & 3.21 & 1.83 & 3.07 & 3.75 & 2.14 \\
 Total & 1.64 & 3.10 & 2.06 & 2.26 & 3.83 & 2.55 \\
 SrAlGe &&&&&&\\
 Total & 1.38 & 3.70 & 2.90 & 1.62 & 4.36 & 3.42 \\
\end{tabular}
\end{ruledtabular}
\end{table}

The character of wave functions can be found from the total density of states(DOS) and partial DOS in Fig. \ref{fig4}.
The total density of states at the Fermi level is 1.64 states/eV which is higher than that of MgB$_2$(0.72 states/eV).\cite{kortus}
The higher DOS of BaAlGe compared to MgB$_2$ is due to Ba $d$ orbitals as shown in Fig. \ref{fig4}(d).
An energy gap of 0.3 eV is explcitly observed between $\pi$ and $\pi^*$ bands around -0.7 eV, which can also be seen in the band plot of Fig. \ref{fig2}.
The sharp peaks at -1.6 eV and -0.28 eV originate mostly from Ge p$_z$ and Al p$_z$ orbitals, respectively.
Figure \ref{fig4}(d) shows that Sr $d$ orbitals contribute considerably to the DOS at the Fermi level, like other ternary silicides;
Sr $d_{xy}$ contributes more than Sr $d_{z^2}$ at the Fermi level, as shown in Fig. \ref{fig4}(d).

\subsection{SrAlGe}
\begin{figure}[bt]
\epsfxsize=\hsize \center{\epsfbox{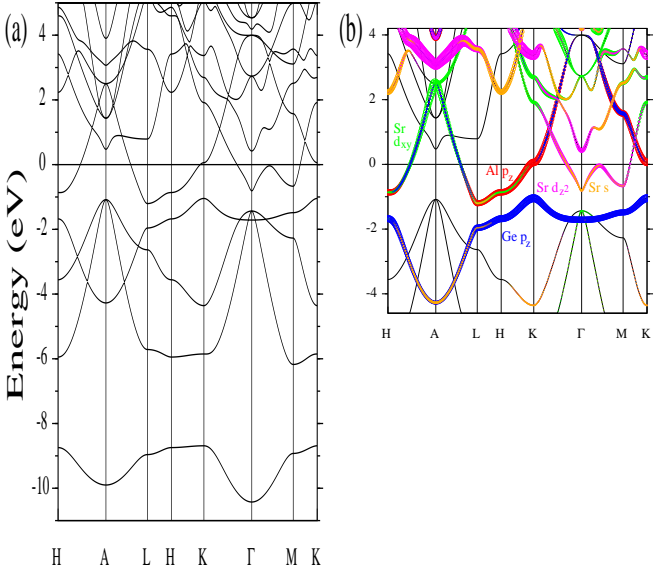}}
\caption{(a) Energy bands and (b) major orbital components of SrAlGe around $E_F$(set to 0).
 \label{fig5}}
\end{figure}

\begin{figure}[bt]
\centering
\begin{tabular}{cc}
\epsfig{file=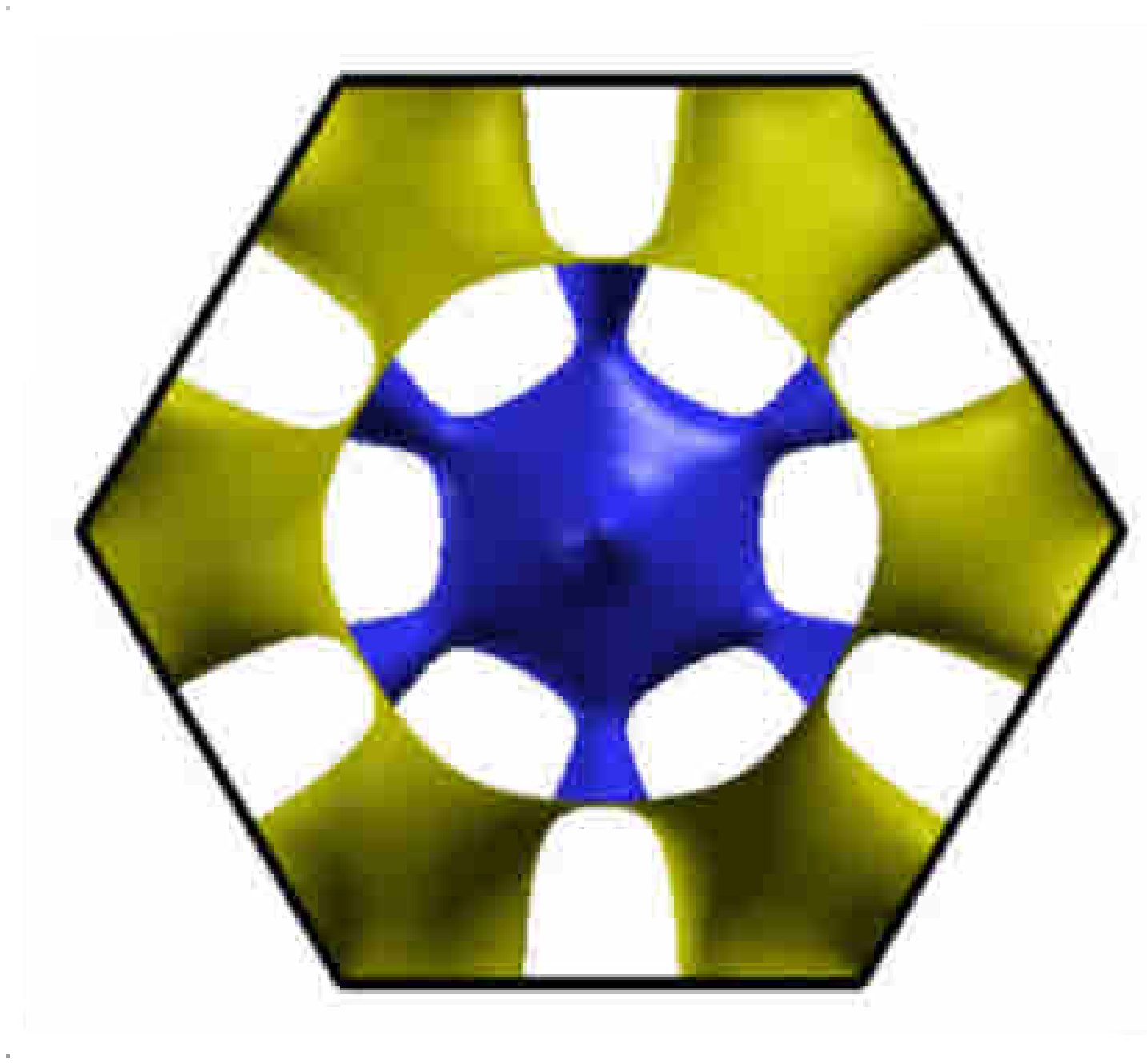,width=0.5\linewidth,clip=} &
\epsfig{file=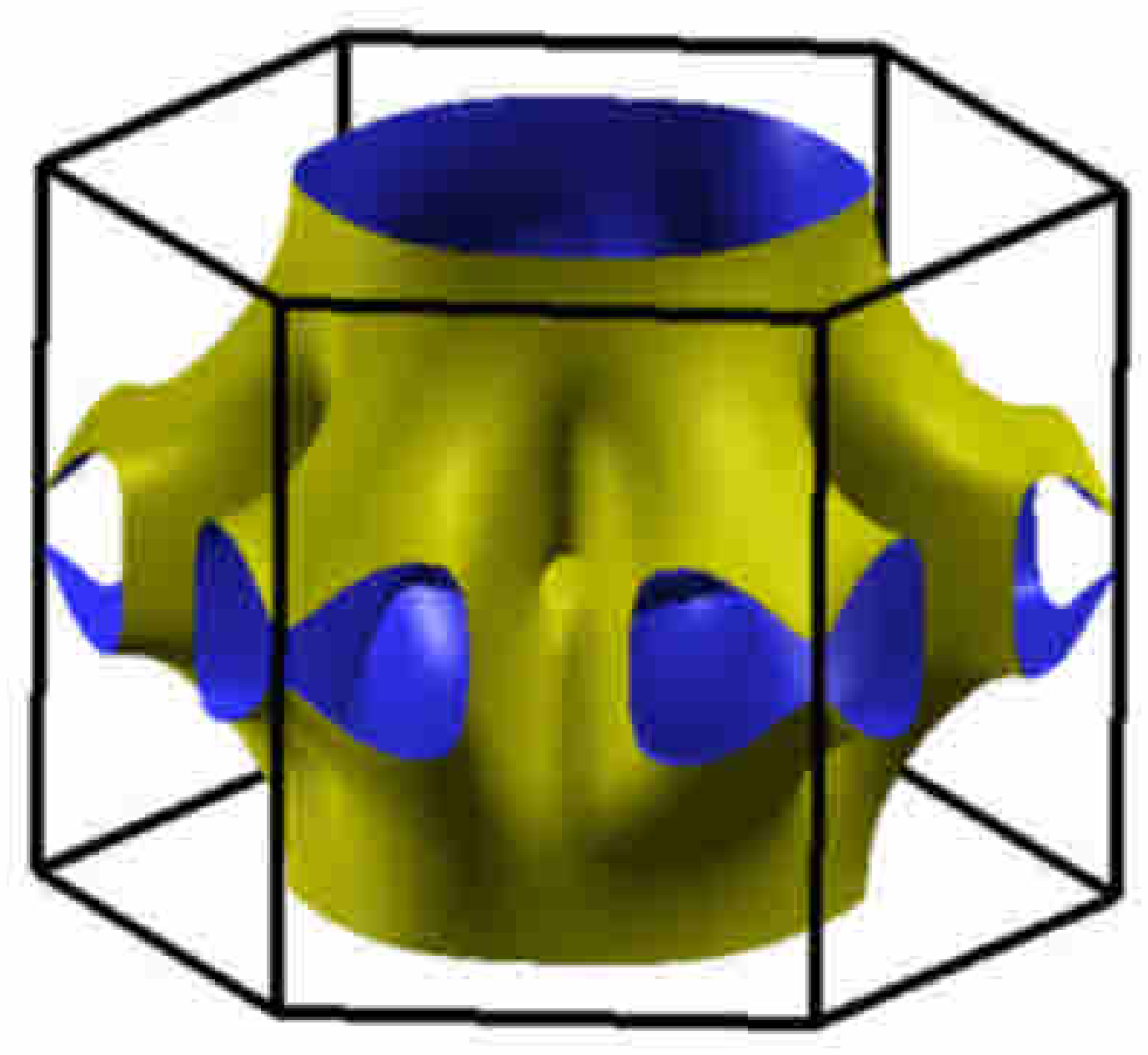,width=0.5\linewidth,clip=}\\
\mbox{\bf (a)} & \mbox{\bf (b)}\\
\epsfig{file=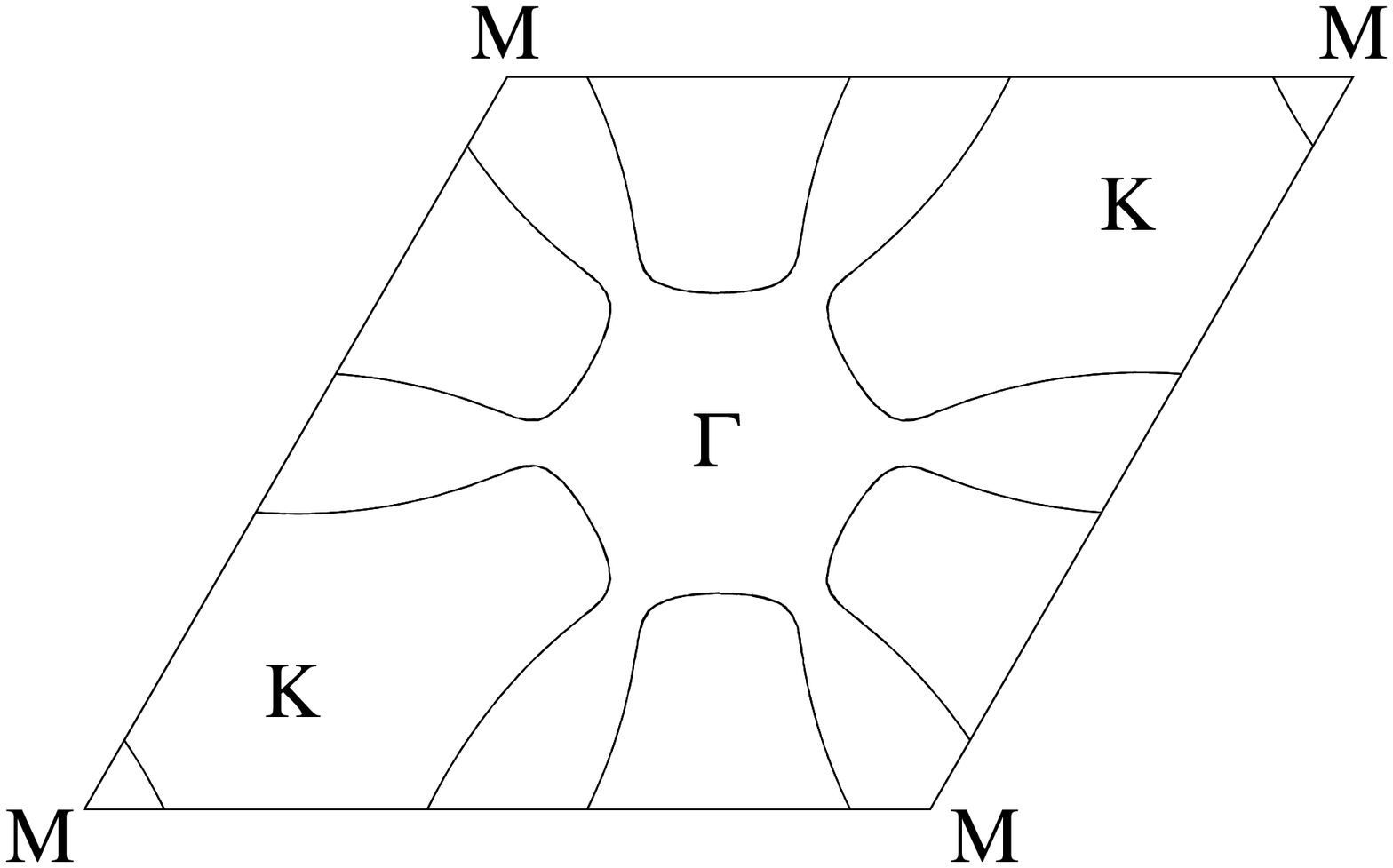,width=0.5\linewidth,clip=} &
\epsfig{file=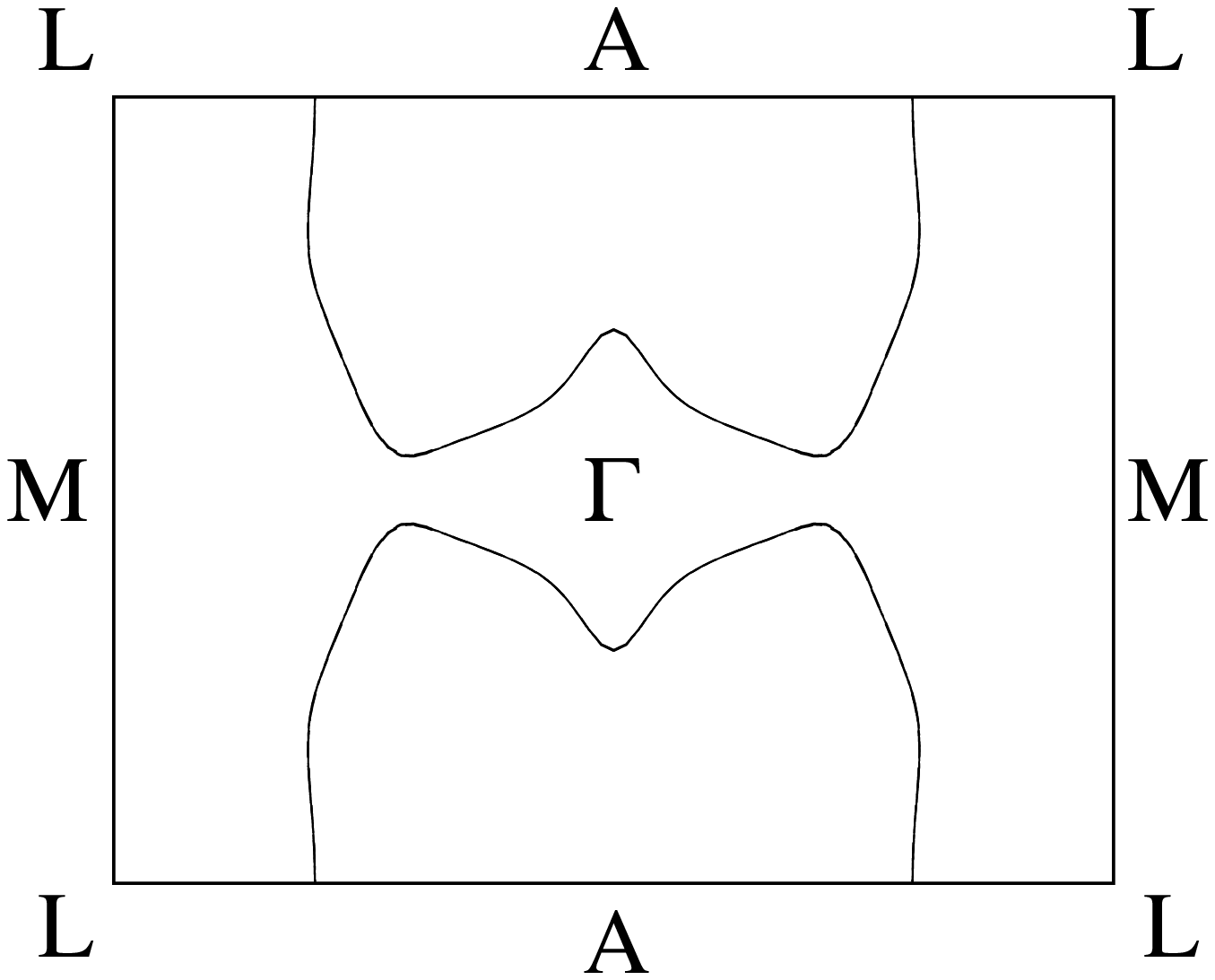,width=0.4\linewidth,clip=}\\
\mbox{\bf (c)} & \mbox{\bf (d)}
\end{tabular}
\caption{Fermi surfaces of SrAlGe. (a) Top view and (b) side view. In (b), the body center of hexagonal prism is the $\Gamma$ point. Blue and golden colors represent electron-like and hole-like surfaces, respectively. Cross-section in (c)(001) plane and (d)(100) plane.\label{fig6}}
\end{figure}

The $\sigma$ band in Fig. \ref{fig5} between about -6 eV and -1 eV is the hybridized band between Al $s,p_{x,y}$, and Ge $p_{x,y}$ orbitals.
The width of the $\sigma$ band is 5.1 eV where the minimum and maximum are located at $M$ and $K$, respectively,
which is a little bit wider than that of BaAlGe.
The little difference of band width between the two materials can be attributed to their similar in-plane lattice constants.
The $\pi^*$ band has a band width of 5.2 eV or 0.8 eV wider than that of BaAlGe,
which is attributed to a smaller $c/a$ ratio, 1.10, of SrAlGe compared to 1.18 of BaAlGe (1.14 for MgB$_2$).
The fifth band along $HK$ shows the wider dispersion of 0.92 eV than that of BaAlGe (0.10 eV) which is due to more hybridization of the Al $p_z$ orbitals along the $z$ direction.
The strong hybridization of the Al $p_z$ orbital enhances the electron-phonon coupling along the $z$ direction which plays an important role in the superconductivity of SrAlGe.\cite{kuroiwa2008}
Note that the fifth band is below the Fermi level along $\Gamma M$ which gives a connected Fermi surface in Fig. \ref{fig6}.
Band components around the Fermi level are displayed in Fig.\ref{fig5}(b).
The $\pi$ band of mostly Ge $p_z$ origin is also fully occupied in SrAlGe as in BaAlGe.

The topology of the Fermi surface of SrAlGe in Fig.\ref{fig6} is in sharp contrast to that of BaAlGe.
There is only one Fermi surface, since the inner and outer Fermi surfaces are connected.
The circular hole around the $K$ point in BaAlGe does not appear in SrAlGe.
The cylindrical 2D character decreases in SrAlGe compared to BaAlGe and shows more 3D character: compare cylindrical Fermi surfaces in Fig. \ref{fig3}(d) and Fig. \ref{fig6}(d).
In-plane and out-of-plane Fermi velocities are
$\langle v_{x,y}^2\rangle^{1/2}$=3.70$\times 10^7$ cm/s and $\langle v_z^2\rangle^{1/2}$=2.90$\times 10^7$ cm/s, respectively.
The plasma frequencies are $\Omega_{x,y}$=4.36 eV and $\Omega_z$=3.42 eV.
Both the Fermi velocities and the plasma frequencies of SrAlGe are larger than those of BaAlGe in both directions as shown in Tab. \ref{table1} which are characteristic of more delocalized bands.
The anisotropic conductivity ratio, 1.62, is smaller than for BaAlGe which is also a feature of the more delocalized band of SrAlGe; however, it is higher than for MgB$_2$.
Since the two Fermi surfaces are connected, contributions from each Fermi surface are not calculated.
The connected Fermi surface may give rise to a higher $T_c$ of SrAlGe than BaAlGe since both Fermi surfaces can contribute to superconductivity with one superconducing energy gap.
Thus, two factors contribute for the higher $T_c$ of SrAlGe than BaAlGe: the enhanced electron-phonon coupling along the $z$ direction due to the Al $p_z$ orbital and the connected Fermi surface.

\begin{figure}[bt]
\epsfxsize=0.7\hsize \center{\epsfbox{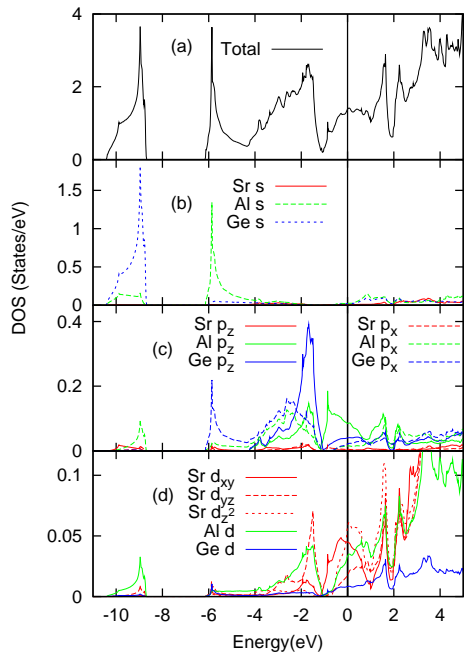}}
\caption{Total and partial density of states of SrAlGe.\label{fig7}}
\end{figure}

The total and partial DOS are displayed in Fig. \ref{fig7};
the total density of states at $E_F$ is 1.38 states/eV which is smaller than that of BaAlGe(1.64 states/eV).
The lower total DOS is mainly due to the wide dispersion of the Al $p_z$ orbital.
There is no gap in the DOS between $\pi$ and $\pi^*$ in Fig. \ref{fig7}(c) while two bands are split in the band plot of Fig. \ref{fig5}.
There are sharp peaks at -1.6 eV and -0.28eV which come from Ge p$_z$ and Al p$_z$ orbitals, respectively.
Figure \ref{fig7}(d) shows that Sr $d$ orbitals in SrAlGe play an important role at the Fermi surface like BaAlGe.
However, the contribution from each orbital is reversed.
Sr $d_{z^2}$ contributes more than Sr $d_{xy}$ at $E_F$ as shown in Fig. \ref{fig7}(d), which is consistent
with the larger dispersion of the $\pi^*$ band in SrAlGe than in BaAlGe.

\section{Summary and Conclusion}
The electronic structure of BaAlGe is determined to be similar to that of ternary silicides with two separate Fermi surfaces;
the $\pi^*$ band is partially occupied to show metallic behavior while the $\pi$ band of Ge $p_z$ origin is fully occupied.
SrAlGe has a wider band width than BaAlGe.
Interlayer coupling is made by non-bonding orbitals such as metal(Ba, Sr) $s, d_{z^2}$ and Al $s$, Ge $s$ orbitals.
The Fermi velocity in the $z$ direction is faster than in the $x$ direction on the pocket Fermi surface of BaAlGe,
which is due to inter-layer coupling.
The topology of the two materials is very different: There are two Fermi surfaces in BaAlGe while they are connected in  SrAlGe.
A large cylindrical sheet of the Fermi surface originates from antibonding $\pi^*$ orbitals, while the electron pocket comes from nonbonding orbitals.
The anisotropy ratio decreases from BaAlGe to SrAlGe which is due to the greater 3D character of SrAlGe.
The higher $T_c$ of SrAlGe than BaAlGe is attributed to both the strong electron-phonon coupling along the $z$ direction and the connected Fermi surfaces.

\begin{acknowledgments}
  SJY acknowledges the sabbatical research Grant by Gyeongsang National University.
  AJF is supported by the Department of Energy (DE-FG02-88ER45382).
\end{acknowledgments}
\bibliography{sag}

\providecommand{\noopsort}[1]{}\providecommand{\singleletter}[1]{#1}%
\begin{thebibliography}{29}%
\makeatletter
\providecommand \@ifxundefined [1]{%
 \@ifx{#1\undefined}
}%
\providecommand \@ifnum [1]{%
 \ifnum #1\expandafter \@firstoftwo
 \else \expandafter \@secondoftwo
 \fi
}%
\providecommand \@ifx [1]{%
 \ifx #1\expandafter \@firstoftwo
 \else \expandafter \@secondoftwo
 \fi
}%
\providecommand \natexlab [1]{#1}%
\providecommand \enquote  [1]{``#1''}%
\providecommand \bibnamefont  [1]{#1}%
\providecommand \bibfnamefont [1]{#1}%
\providecommand \citenamefont [1]{#1}%
\providecommand \href@noop [0]{\@secondoftwo}%
\providecommand \href [0]{\begingroup \@sanitize@url \@href}%
\providecommand \@href[1]{\@@startlink{#1}\@@href}%
\providecommand \@@href[1]{\endgroup#1\@@endlink}%
\providecommand \@sanitize@url [0]{\catcode `\\12\catcode `\$12\catcode
  `\&12\catcode `\#12\catcode `\^12\catcode `\_12\catcode `\%12\relax}%
\providecommand \@@startlink[1]{}%
\providecommand \@@endlink[0]{}%
\providecommand \url  [0]{\begingroup\@sanitize@url \@url }%
\providecommand \@url [1]{\endgroup\@href {#1}{\urlprefix }}%
\providecommand \urlprefix  [0]{URL }%
\providecommand \Eprint [0]{\href }%
\providecommand \doibase [0]{http://dx.doi.org/}%
\providecommand \selectlanguage [0]{\@gobble}%
\providecommand \bibinfo  [0]{\@secondoftwo}%
\providecommand \bibfield  [0]{\@secondoftwo}%
\providecommand \translation [1]{[#1]}%
\providecommand \BibitemOpen [0]{}%
\providecommand \bibitemStop [0]{}%
\providecommand \bibitemNoStop [0]{.\EOS\space}%
\providecommand \EOS [0]{\spacefactor3000\relax}%
\providecommand \BibitemShut  [1]{\csname bibitem#1\endcsname}%
\let\auto@bib@innerbib\@empty
\bibitem [{\citenamefont {Nagamatsu}\ \emph {et~al.}(2001)\citenamefont
  {Nagamatsu}, \citenamefont {Nakagawa}, \citenamefont {Muranaka},
  \citenamefont {Zenitani},\ and\ \citenamefont {Akimitsu}}]{nagamatsu}%
  \BibitemOpen
  \bibfield  {author} {\bibinfo {author} {\bibfnamefont {J.}~\bibnamefont
  {Nagamatsu}}, \bibinfo {author} {\bibfnamefont {N.}~\bibnamefont {Nakagawa}},
  \bibinfo {author} {\bibfnamefont {T.}~\bibnamefont {Muranaka}}, \bibinfo
  {author} {\bibfnamefont {Y.}~\bibnamefont {Zenitani}}, \ and\ \bibinfo
  {author} {\bibfnamefont {J.}~\bibnamefont {Akimitsu}},\ }\href@noop {}
  {\bibfield  {journal} {\bibinfo  {journal} {Nature (London)}\ }\textbf
  {\bibinfo {volume} {410}},\ \bibinfo {pages} {63} (\bibinfo {year}
  {2001})}\BibitemShut {NoStop}%
\bibitem [{\citenamefont {Moshchalkov}\ \emph {et~al.}(2009)\citenamefont
  {Moshchalkov}, \citenamefont {Menghini}, \citenamefont {Nishio},
  \citenamefont {Chen}, \citenamefont {Silhanek}, \citenamefont {Dao},
  \citenamefont {Chibotaru}, \citenamefont {Zhigadlo},\ and\ \citenamefont
  {Karpinski}}]{moshchalkov}%
  \BibitemOpen
  \bibfield  {author} {\bibinfo {author} {\bibfnamefont {V.}~\bibnamefont
  {Moshchalkov}}, \bibinfo {author} {\bibfnamefont {M.}~\bibnamefont
  {Menghini}}, \bibinfo {author} {\bibfnamefont {T.}~\bibnamefont {Nishio}},
  \bibinfo {author} {\bibfnamefont {Q.~H.}\ \bibnamefont {Chen}}, \bibinfo
  {author} {\bibfnamefont {A.~V.}\ \bibnamefont {Silhanek}}, \bibinfo {author}
  {\bibfnamefont {V.~H.}\ \bibnamefont {Dao}}, \bibinfo {author} {\bibfnamefont
  {L.~F.}\ \bibnamefont {Chibotaru}}, \bibinfo {author} {\bibfnamefont {N.~D.}\
  \bibnamefont {Zhigadlo}}, \ and\ \bibinfo {author} {\bibfnamefont
  {J.}~\bibnamefont {Karpinski}},\ }\href@noop {} {\bibfield  {journal}
  {\bibinfo  {journal} {Phys. Rev. Lett.}\ }\textbf {\bibinfo {volume} {102}},\
  \bibinfo {pages} {117001} (\bibinfo {year} {2009})}\BibitemShut {NoStop}%
\bibitem [{\citenamefont {Collings}\ \emph {et~al.}(2008)\citenamefont
  {Collings}, \citenamefont {Sumption}, \citenamefont {Bhatia}, \citenamefont
  {Susner},\ and\ \citenamefont {Bohnenstiehl}}]{collings}%
  \BibitemOpen
  \bibfield  {author} {\bibinfo {author} {\bibfnamefont {E.~W.}\ \bibnamefont
  {Collings}}, \bibinfo {author} {\bibfnamefont {M.~D.}\ \bibnamefont
  {Sumption}}, \bibinfo {author} {\bibfnamefont {M.}~\bibnamefont {Bhatia}},
  \bibinfo {author} {\bibfnamefont {M.~A.}\ \bibnamefont {Susner}}, \ and\
  \bibinfo {author} {\bibfnamefont {S.~D.}\ \bibnamefont {Bohnenstiehl}},\
  }\href@noop {} {\bibfield  {journal} {\bibinfo  {journal} {Supercond. Sci.
  Technol.}\ }\textbf {\bibinfo {volume} {21}},\ \bibinfo {pages} {103001}
  (\bibinfo {year} {2008})}\BibitemShut {NoStop}%
\bibitem [{\citenamefont {Hoffmann}\ and\ \citenamefont
  {ottgen}(2001)}]{hoffmann}%
  \BibitemOpen
  \bibfield  {author} {\bibinfo {author} {\bibfnamefont {R.-D.}\ \bibnamefont
  {Hoffmann}}\ and\ \bibinfo {author} {\bibfnamefont {R.~P.}\ \bibnamefont
  {ottgen}},\ }\href@noop {} {\bibfield  {journal} {\bibinfo  {journal} {Z.
  Kristallogra.}\ }\textbf {\bibinfo {volume} {216}},\ \bibinfo {pages} {127}
  (\bibinfo {year} {2001})}\BibitemShut {NoStop}%
\bibitem [{\citenamefont {An}\ and\ \citenamefont {Pickett}(2001)}]{an}%
  \BibitemOpen
  \bibfield  {author} {\bibinfo {author} {\bibfnamefont {J.~M.}\ \bibnamefont
  {An}}\ and\ \bibinfo {author} {\bibfnamefont {W.~E.}\ \bibnamefont
  {Pickett}},\ }\href@noop {} {\bibfield  {journal} {\bibinfo  {journal} {Phys.
  Rev. Lett.}\ }\textbf {\bibinfo {volume} {86}},\ \bibinfo {pages} {4366}
  (\bibinfo {year} {2001})}\BibitemShut {NoStop}%
\bibitem [{\citenamefont {Kortus}\ \emph {et~al.}(2001)\citenamefont {Kortus},
  \citenamefont {Mazin}, \citenamefont {Belashchenko}, \citenamefont
  {Antropov},\ and\ \citenamefont {Boyer}}]{kortus}%
  \BibitemOpen
  \bibfield  {author} {\bibinfo {author} {\bibfnamefont {J.}~\bibnamefont
  {Kortus}}, \bibinfo {author} {\bibfnamefont {I.~I.}\ \bibnamefont {Mazin}},
  \bibinfo {author} {\bibfnamefont {K.~D.}\ \bibnamefont {Belashchenko}},
  \bibinfo {author} {\bibfnamefont {V.~P.}\ \bibnamefont {Antropov}}, \ and\
  \bibinfo {author} {\bibfnamefont {L.~L.}\ \bibnamefont {Boyer}},\ }\href@noop
  {} {\bibfield  {journal} {\bibinfo  {journal} {Phys. Rev. Lett.}\ }\textbf
  {\bibinfo {volume} {86}},\ \bibinfo {pages} {4656} (\bibinfo {year}
  {2001})}\BibitemShut {NoStop}%
\bibitem [{\citenamefont {Choi}\ \emph {et~al.}(2002)\citenamefont {Choi},
  \citenamefont {Roundy}, \citenamefont {Sun}, \citenamefont {Cohen},\ and\
  \citenamefont {Louie}}]{choi}%
  \BibitemOpen
  \bibfield  {author} {\bibinfo {author} {\bibfnamefont {H.~J.}\ \bibnamefont
  {Choi}}, \bibinfo {author} {\bibfnamefont {D.}~\bibnamefont {Roundy}},
  \bibinfo {author} {\bibfnamefont {H.}~\bibnamefont {Sun}}, \bibinfo {author}
  {\bibfnamefont {M.~L.}\ \bibnamefont {Cohen}}, \ and\ \bibinfo {author}
  {\bibfnamefont {S.~G.}\ \bibnamefont {Louie}},\ }\href@noop {} {\bibfield
  {journal} {\bibinfo  {journal} {Nature (London)}\ }\textbf {\bibinfo {volume}
  {418}},\ \bibinfo {pages} {758} (\bibinfo {year} {2002})}\BibitemShut
  {NoStop}%
\bibitem [{\citenamefont {Sanfilippo}\ \emph {et~al.}(2000)\citenamefont
  {Sanfilippo}, \citenamefont {Elsinger}, \citenamefont {N\'u\~nez Regueiro},
  \citenamefont {Laborde}, \citenamefont {LeFloch}, \citenamefont {Affronte},
  \citenamefont {Olcese},\ and\ \citenamefont {Palenzona}}]{sanfilippo}%
  \BibitemOpen
  \bibfield  {author} {\bibinfo {author} {\bibfnamefont {S.}~\bibnamefont
  {Sanfilippo}}, \bibinfo {author} {\bibfnamefont {H.}~\bibnamefont
  {Elsinger}}, \bibinfo {author} {\bibfnamefont {M.}~\bibnamefont {N\'u\~nez
  Regueiro}}, \bibinfo {author} {\bibfnamefont {O.}~\bibnamefont {Laborde}},
  \bibinfo {author} {\bibfnamefont {S.}~\bibnamefont {LeFloch}}, \bibinfo
  {author} {\bibfnamefont {M.}~\bibnamefont {Affronte}}, \bibinfo {author}
  {\bibfnamefont {G.~L.}\ \bibnamefont {Olcese}}, \ and\ \bibinfo {author}
  {\bibfnamefont {A.}~\bibnamefont {Palenzona}},\ }\href@noop {} {\bibfield
  {journal} {\bibinfo  {journal} {Phys. Rev. B}\ }\textbf {\bibinfo {volume}
  {61}},\ \bibinfo {pages} {R3800} (\bibinfo {year} {2000})}\BibitemShut
  {NoStop}%
\bibitem [{\citenamefont {Fahy}\ and\ \citenamefont {Hamann}(1990)}]{fahy}%
  \BibitemOpen
  \bibfield  {author} {\bibinfo {author} {\bibfnamefont {S.}~\bibnamefont
  {Fahy}}\ and\ \bibinfo {author} {\bibfnamefont {D.~R.}\ \bibnamefont
  {Hamann}},\ }\href@noop {} {\bibfield  {journal} {\bibinfo  {journal} {Phys.
  Rev. B}\ }\textbf {\bibinfo {volume} {41}},\ \bibinfo {pages} {7587}
  (\bibinfo {year} {1990})}\BibitemShut {NoStop}%
\bibitem [{\citenamefont {Imai}\ \emph {et~al.}(2002)\citenamefont {Imai},
  \citenamefont {Nishida}, \citenamefont {Kimura},\ and\ \citenamefont
  {Abe}}]{imai2002}%
  \BibitemOpen
  \bibfield  {author} {\bibinfo {author} {\bibfnamefont {M.}~\bibnamefont
  {Imai}}, \bibinfo {author} {\bibfnamefont {K.}~\bibnamefont {Nishida}},
  \bibinfo {author} {\bibfnamefont {T.}~\bibnamefont {Kimura}}, \ and\ \bibinfo
  {author} {\bibfnamefont {H.}~\bibnamefont {Abe}},\ }\href@noop {} {\bibfield
  {journal} {\bibinfo  {journal} {Appl. Phys. Lett.}\ }\textbf {\bibinfo
  {volume} {80}},\ \bibinfo {pages} {1019} (\bibinfo {year}
  {2002})}\BibitemShut {NoStop}%
\bibitem [{\citenamefont {Kuroiwa}\ \emph {et~al.}(2006)\citenamefont
  {Kuroiwa}, \citenamefont {Sagayama}, \citenamefont {Kakiuchi}, \citenamefont
  {Sawa}, \citenamefont {Noda},\ and\ \citenamefont {Akimitsu}}]{kuroiwa2006}%
  \BibitemOpen
  \bibfield  {author} {\bibinfo {author} {\bibfnamefont {S.}~\bibnamefont
  {Kuroiwa}}, \bibinfo {author} {\bibfnamefont {H.}~\bibnamefont {Sagayama}},
  \bibinfo {author} {\bibfnamefont {T.}~\bibnamefont {Kakiuchi}}, \bibinfo
  {author} {\bibfnamefont {H.}~\bibnamefont {Sawa}}, \bibinfo {author}
  {\bibfnamefont {Y.}~\bibnamefont {Noda}}, \ and\ \bibinfo {author}
  {\bibfnamefont {J.}~\bibnamefont {Akimitsu}},\ }\href@noop {} {\bibfield
  {journal} {\bibinfo  {journal} {Phys. Rev. B}\ }\textbf {\bibinfo {volume}
  {74}},\ \bibinfo {pages} {014517} (\bibinfo {year} {2006})}\BibitemShut
  {NoStop}%
\bibitem [{\citenamefont {Huang}\ \emph {et~al.}(2004)\citenamefont {Huang},
  \citenamefont {Chen}, \citenamefont {Liu},\ and\ \citenamefont
  {Xing}}]{huang}%
  \BibitemOpen
  \bibfield  {author} {\bibinfo {author} {\bibfnamefont {G.~Q.}\ \bibnamefont
  {Huang}}, \bibinfo {author} {\bibfnamefont {L.~F.}\ \bibnamefont {Chen}},
  \bibinfo {author} {\bibfnamefont {M.}~\bibnamefont {Liu}}, \ and\ \bibinfo
  {author} {\bibfnamefont {D.~Y.}\ \bibnamefont {Xing}},\ }\href@noop {}
  {\bibfield  {journal} {\bibinfo  {journal} {Phys. Rev. B}\ }\textbf {\bibinfo
  {volume} {69}},\ \bibinfo {pages} {064509} (\bibinfo {year}
  {2004})}\BibitemShut {NoStop}%
\bibitem [{\citenamefont {Lorenz}\ \emph {et~al.}(2002)\citenamefont {Lorenz},
  \citenamefont {Lenzi}, \citenamefont {Cmaidalka}, \citenamefont {Meng},
  \citenamefont {Sun}, \citenamefont {Xue},\ and\ \citenamefont
  {Chu}}]{lorenz}%
  \BibitemOpen
  \bibfield  {author} {\bibinfo {author} {\bibfnamefont {B.}~\bibnamefont
  {Lorenz}}, \bibinfo {author} {\bibfnamefont {J.}~\bibnamefont {Lenzi}},
  \bibinfo {author} {\bibfnamefont {J.}~\bibnamefont {Cmaidalka}}, \bibinfo
  {author} {\bibfnamefont {R.~L.}\ \bibnamefont {Meng}}, \bibinfo {author}
  {\bibfnamefont {Y.~Y.}\ \bibnamefont {Sun}}, \bibinfo {author} {\bibfnamefont
  {Y.~Y.}\ \bibnamefont {Xue}}, \ and\ \bibinfo {author} {\bibfnamefont
  {C.~W.}\ \bibnamefont {Chu}},\ }\href@noop {} {\bibfield  {journal} {\bibinfo
   {journal} {Physica C}\ }\textbf {\bibinfo {volume} {383}},\ \bibinfo {pages}
  {191} (\bibinfo {year} {2002})}\BibitemShut {NoStop}%
\bibitem [{\citenamefont {Mazin}\ and\ \citenamefont
  {Papaconstantopoulos}(2004)}]{mazin}%
  \BibitemOpen
  \bibfield  {author} {\bibinfo {author} {\bibfnamefont {I.~I.}\ \bibnamefont
  {Mazin}}\ and\ \bibinfo {author} {\bibfnamefont {D.~A.}\ \bibnamefont
  {Papaconstantopoulos}},\ }\href@noop {} {\bibfield  {journal} {\bibinfo
  {journal} {Phys. Rev. B}\ }\textbf {\bibinfo {volume} {69}},\ \bibinfo
  {pages} {180512} (\bibinfo {year} {2004})}\BibitemShut {NoStop}%
\bibitem [{\citenamefont {Giantomassi}\ \emph {et~al.}(2005)\citenamefont
  {Giantomassi}, \citenamefont {Boeri},\ and\ \citenamefont
  {Bachelet}}]{giantomassi}%
  \BibitemOpen
  \bibfield  {author} {\bibinfo {author} {\bibfnamefont {M.}~\bibnamefont
  {Giantomassi}}, \bibinfo {author} {\bibfnamefont {L.}~\bibnamefont {Boeri}},
  \ and\ \bibinfo {author} {\bibfnamefont {G.~B.}\ \bibnamefont {Bachelet}},\
  }\href@noop {} {\bibfield  {journal} {\bibinfo  {journal} {Phys. Rev. B}\
  }\textbf {\bibinfo {volume} {72}},\ \bibinfo {pages} {224512} (\bibinfo
  {year} {2005})}\BibitemShut {NoStop}%
\bibitem [{\citenamefont {Kuroiwa}\ \emph {et~al.}(2008)\citenamefont
  {Kuroiwa}, \citenamefont {Baron}, \citenamefont {Muranaka}, \citenamefont
  {Heid}, \citenamefont {Bohnen},\ and\ \citenamefont
  {Akimitsu}}]{kuroiwa2008}%
  \BibitemOpen
  \bibfield  {author} {\bibinfo {author} {\bibfnamefont {S.}~\bibnamefont
  {Kuroiwa}}, \bibinfo {author} {\bibfnamefont {A.~Q.~R.}\ \bibnamefont
  {Baron}}, \bibinfo {author} {\bibfnamefont {T.}~\bibnamefont {Muranaka}},
  \bibinfo {author} {\bibfnamefont {R.}~\bibnamefont {Heid}}, \bibinfo {author}
  {\bibfnamefont {K.-P.}\ \bibnamefont {Bohnen}}, \ and\ \bibinfo {author}
  {\bibfnamefont {J.}~\bibnamefont {Akimitsu}},\ }\href@noop {} {\bibfield
  {journal} {\bibinfo  {journal} {Phys. Rev. B}\ }\textbf {\bibinfo {volume}
  {77}},\ \bibinfo {pages} {140503} (\bibinfo {year} {2008})}\BibitemShut
  {NoStop}%
\bibitem [{\citenamefont {Evans}\ \emph {et~al.}(2009)\citenamefont {Evans},
  \citenamefont {Wu}, \citenamefont {Kranak}, \citenamefont {Newman},
  \citenamefont {Reller}, \citenamefont {Garcia-Garcia},\ and\ \citenamefont
  {aussermann}}]{evans}%
  \BibitemOpen
  \bibfield  {author} {\bibinfo {author} {\bibfnamefont {M.~J.}\ \bibnamefont
  {Evans}}, \bibinfo {author} {\bibfnamefont {Y.}~\bibnamefont {Wu}}, \bibinfo
  {author} {\bibfnamefont {V.~F.}\ \bibnamefont {Kranak}}, \bibinfo {author}
  {\bibfnamefont {N.}~\bibnamefont {Newman}}, \bibinfo {author} {\bibfnamefont
  {A.}~\bibnamefont {Reller}}, \bibinfo {author} {\bibfnamefont {F.~J.}\
  \bibnamefont {Garcia-Garcia}}, \ and\ \bibinfo {author} {\bibfnamefont
  {U.~H.}\ \bibnamefont {aussermann}},\ }\href@noop {} {\bibfield  {journal}
  {\bibinfo  {journal} {Phys. Rev. B}\ }\textbf {\bibinfo {volume} {80}},\
  \bibinfo {pages} {064514} (\bibinfo {year} {2009})}\BibitemShut {NoStop}%
\bibitem [{\citenamefont {Wimmer}\ \emph {et~al.}(1981)\citenamefont {Wimmer},
  \citenamefont {Krakauer}, \citenamefont {Weinert},\ and\ \citenamefont
  {Freeman}}]{flapw}%
  \BibitemOpen
  \bibfield  {author} {\bibinfo {author} {\bibfnamefont {E.}~\bibnamefont
  {Wimmer}}, \bibinfo {author} {\bibfnamefont {H.}~\bibnamefont {Krakauer}},
  \bibinfo {author} {\bibfnamefont {M.}~\bibnamefont {Weinert}}, \ and\
  \bibinfo {author} {\bibfnamefont {A.~J.}\ \bibnamefont {Freeman}},\
  }\href@noop {} {\bibfield  {journal} {\bibinfo  {journal} {Phys. Rev. B}\
  }\textbf {\bibinfo {volume} {24}},\ \bibinfo {pages} {864} (\bibinfo {year}
  {1981})}\BibitemShut {NoStop}%
\bibitem [{\citenamefont {Weinert}\ \emph {et~al.}(1982)\citenamefont
  {Weinert}, \citenamefont {Wimmer},\ and\ \citenamefont {Freeman}}]{flapw2}%
  \BibitemOpen
  \bibfield  {author} {\bibinfo {author} {\bibfnamefont {M.}~\bibnamefont
  {Weinert}}, \bibinfo {author} {\bibfnamefont {E.}~\bibnamefont {Wimmer}}, \
  and\ \bibinfo {author} {\bibfnamefont {A.~J.}\ \bibnamefont {Freeman}},\
  }\href@noop {} {\bibfield  {journal} {\bibinfo  {journal} {Phys. Rev. B}\
  }\textbf {\bibinfo {volume} {26}},\ \bibinfo {pages} {4571} (\bibinfo {year}
  {1982})}\BibitemShut {NoStop}%
\bibitem [{\citenamefont {Hedin}\ and\ \citenamefont
  {Lundqvist}(1971)}]{hedin}%
  \BibitemOpen
  \bibfield  {author} {\bibinfo {author} {\bibfnamefont {L.}~\bibnamefont
  {Hedin}}\ and\ \bibinfo {author} {\bibfnamefont {B.~I.}\ \bibnamefont
  {Lundqvist}},\ }\href@noop {} {\bibfield  {journal} {\bibinfo  {journal} {J.
  Phys. C}\ }\textbf {\bibinfo {volume} {4}},\ \bibinfo {pages} {2064}
  (\bibinfo {year} {1971})}\BibitemShut {NoStop}%
\bibitem [{\citenamefont {Weinert}\ \emph {et~al.}(2009)\citenamefont
  {Weinert}, \citenamefont {Schneider}, \citenamefont {Podloucky},\ and\
  \citenamefont {redinger}}]{explicit}%
  \BibitemOpen
  \bibfield  {author} {\bibinfo {author} {\bibfnamefont {M.}~\bibnamefont
  {Weinert}}, \bibinfo {author} {\bibfnamefont {G.}~\bibnamefont {Schneider}},
  \bibinfo {author} {\bibfnamefont {R.}~\bibnamefont {Podloucky}}, \ and\
  \bibinfo {author} {\bibfnamefont {J.}~\bibnamefont {redinger}},\ }\href@noop
  {} {\bibfield  {journal} {\bibinfo  {journal} {J. Phys.: Cond. Matt.}\
  }\textbf {\bibinfo {volume} {21}},\ \bibinfo {pages} {084201} (\bibinfo
  {year} {2009})}\BibitemShut {NoStop}%
\bibitem [{\citenamefont {Ramirez}\ and\ \citenamefont
  {Bh\"om}(1988)}]{ramirez}%
  \BibitemOpen
  \bibfield  {author} {\bibinfo {author} {\bibfnamefont {R.}~\bibnamefont
  {Ramirez}}\ and\ \bibinfo {author} {\bibfnamefont {M.~C.}\ \bibnamefont
  {Bh\"om}},\ }\href@noop {} {\bibfield  {journal} {\bibinfo  {journal} {Int.
  J. Quant. Chem.}\ }\textbf {\bibinfo {volume} {34}},\ \bibinfo {pages} {571}
  (\bibinfo {year} {1988})}\BibitemShut {NoStop}%
\bibitem [{\citenamefont {Monkhorst}\ and\ \citenamefont {Pack}(1976)}]{kpts}%
  \BibitemOpen
  \bibfield  {author} {\bibinfo {author} {\bibfnamefont {H.~J.}\ \bibnamefont
  {Monkhorst}}\ and\ \bibinfo {author} {\bibfnamefont {J.~D.}\ \bibnamefont
  {Pack}},\ }\href@noop {} {\bibfield  {journal} {\bibinfo  {journal} {Phys.
  Rev. B}\ }\textbf {\bibinfo {volume} {13}},\ \bibinfo {pages} {5188}
  (\bibinfo {year} {1976})}\BibitemShut {NoStop}%
\bibitem [{\citenamefont {Bl\"ochl}\ \emph {et~al.}(1994)\citenamefont
  {Bl\"ochl}, \citenamefont {Jepsen},\ and\ \citenamefont {Andersen}}]{blochl}%
  \BibitemOpen
  \bibfield  {author} {\bibinfo {author} {\bibfnamefont {P.~E.}\ \bibnamefont
  {Bl\"ochl}}, \bibinfo {author} {\bibfnamefont {O.}~\bibnamefont {Jepsen}}, \
  and\ \bibinfo {author} {\bibfnamefont {O.~K.}\ \bibnamefont {Andersen}},\
  }\href@noop {} {\bibfield  {journal} {\bibinfo  {journal} {Phys. Rev. B}\
  }\textbf {\bibinfo {volume} {49}},\ \bibinfo {pages} {16223} (\bibinfo {year}
  {1994})}\BibitemShut {NoStop}%
\bibitem [{\citenamefont {Pickett}\ \emph {et~al.}(1988)\citenamefont
  {Pickett}, \citenamefont {Krakauer},\ and\ \citenamefont {Allen}}]{spline}%
  \BibitemOpen
  \bibfield  {author} {\bibinfo {author} {\bibfnamefont {W.~E.}\ \bibnamefont
  {Pickett}}, \bibinfo {author} {\bibfnamefont {H.}~\bibnamefont {Krakauer}}, \
  and\ \bibinfo {author} {\bibfnamefont {P.~B.}\ \bibnamefont {Allen}},\
  }\href@noop {} {\bibfield  {journal} {\bibinfo  {journal} {Phys. Rev. B}\
  }\textbf {\bibinfo {volume} {38}},\ \bibinfo {pages} {2721} (\bibinfo {year}
  {1988})}\BibitemShut {NoStop}%
\bibitem [{\citenamefont {Pearson}(1988)}]{pearson}%
  \BibitemOpen
  \bibfield  {author} {\bibinfo {author} {\bibfnamefont {R.~G.}\ \bibnamefont
  {Pearson}},\ }\href@noop {} {\bibfield  {journal} {\bibinfo  {journal}
  {Inorg. Chem.}\ }\textbf {\bibinfo {volume} {27}},\ \bibinfo {pages} {734}
  (\bibinfo {year} {1988})}\BibitemShut {NoStop}%
\bibitem [{\citenamefont {Mattheiss}\ and\ \citenamefont
  {Hamann}(1984)}]{mattheiss}%
  \BibitemOpen
  \bibfield  {author} {\bibinfo {author} {\bibfnamefont {L.~F.}\ \bibnamefont
  {Mattheiss}}\ and\ \bibinfo {author} {\bibfnamefont {D.~R.}\ \bibnamefont
  {Hamann}},\ }\href@noop {} {\bibfield  {journal} {\bibinfo  {journal} {Phys.
  Rev. B}\ }\textbf {\bibinfo {volume} {30}},\ \bibinfo {pages} {1731}
  (\bibinfo {year} {1984})}\BibitemShut {NoStop}%
\bibitem [{\citenamefont {Kuroiwa}\ \emph {et~al.}(2007)\citenamefont
  {Kuroiwa}, \citenamefont {Nakashima}, \citenamefont {Miyahara}, \citenamefont
  {Furukawa},\ and\ \citenamefont {Akimitsu}}]{kuroiwa2007}%
  \BibitemOpen
  \bibfield  {author} {\bibinfo {author} {\bibfnamefont {S.}~\bibnamefont
  {Kuroiwa}}, \bibinfo {author} {\bibfnamefont {A.}~\bibnamefont {Nakashima}},
  \bibinfo {author} {\bibfnamefont {S.}~\bibnamefont {Miyahara}}, \bibinfo
  {author} {\bibfnamefont {N.}~\bibnamefont {Furukawa}}, \ and\ \bibinfo
  {author} {\bibfnamefont {J.}~\bibnamefont {Akimitsu}},\ }\href@noop {}
  {\bibfield  {journal} {\bibinfo  {journal} {J. Phys. Soc. Jap.}\ }\textbf
  {\bibinfo {volume} {76}},\ \bibinfo {pages} {113705} (\bibinfo {year}
  {2007})}\BibitemShut {NoStop}%
\bibitem [{\citenamefont {Lupi}\ \emph {et~al.}(2008)\citenamefont {Lupi},
  \citenamefont {Baldassarre}, \citenamefont {Ortolani}, \citenamefont {Mirri},
  \citenamefont {Schade}, \citenamefont {Sopracase}, \citenamefont {Tamegai},
  \citenamefont {Fittipaldi}, \citenamefont {Vecchione},\ and\ \citenamefont
  {Calvani}}]{lupi}%
  \BibitemOpen
  \bibfield  {author} {\bibinfo {author} {\bibfnamefont {S.}~\bibnamefont
  {Lupi}}, \bibinfo {author} {\bibfnamefont {L.}~\bibnamefont {Baldassarre}},
  \bibinfo {author} {\bibfnamefont {M.}~\bibnamefont {Ortolani}}, \bibinfo
  {author} {\bibfnamefont {C.}~\bibnamefont {Mirri}}, \bibinfo {author}
  {\bibfnamefont {U.}~\bibnamefont {Schade}}, \bibinfo {author} {\bibfnamefont
  {R.}~\bibnamefont {Sopracase}}, \bibinfo {author} {\bibfnamefont
  {T.}~\bibnamefont {Tamegai}}, \bibinfo {author} {\bibfnamefont
  {R.}~\bibnamefont {Fittipaldi}}, \bibinfo {author} {\bibfnamefont
  {A.}~\bibnamefont {Vecchione}}, \ and\ \bibinfo {author} {\bibfnamefont
  {P.}~\bibnamefont {Calvani}},\ }\href@noop {} {\bibfield  {journal} {\bibinfo
   {journal} {Phys. Rev. B}\ }\textbf {\bibinfo {volume} {77}},\ \bibinfo
  {pages} {054510} (\bibinfo {year} {2008})}\BibitemShut {NoStop}%
\end{thebibliography}%

\end{document}